\documentclass[hyper]{JHEP} 

\usepackage{epsfig}

\newcommand\fverb{\setbox\pippobox=\hbox\bgroup\verb}

\newcommand\fverbdo{\egroup\medskip\noindent%

            \fbox{\unhbox\pippobox}\ }

\newcommand\fverbit{\egroup\item[\fbox{\unhbox\pippobox}]}

\newbox\pippobox

\title{Hamiltonian Analysis of the
Conformal Decomposition of the Gravitational Field}
\author{J. Kluso\v{n}\\
Department of
Theoretical Physics and Astrophysics\\
Faculty of Science, Masaryk University\\
Kotl\'{a}\v{r}sk\'{a} 2, 611 37, Brno\\
Czech Republic\\
E-mail: \email{klu@physics.muni.cz}}
\preprint{}

 \abstract{This short note is devoted to the
  Hamiltonian formulation  of the
 conformal decomposition of the gravitational field that was performed in
 [gr-qc/0501092]. We also analyze the gauge fixed form
 of the theory when we fix the conformal symmetry by imposing the condition
 $\sqrt{g}=1$. }

\def\bD{\mathbf{D}}

\def\tbT{\tilde{\mathbf{T}}}

\def\tpi{\tilde{\pi}}
\def\tmH{\tilde{\mathcal{H}}}

\def\tmG{\tilde{\mG}}
\def\be{\begin{equation}}
\def\bD{\mathbf{D}}
\def\ee{\end{equation}}

\def\bea{\begin{eqnarray}}

\def\eea{\end{eqnarray}}

\def\mH{\mathcal{H}}

\def\bz{\mathbf{z}}

\def\bx{\mathbf{x}}
\def\by{\mathbf{y}}

\newcommand{\hg}{\hat{g}}

\newcommand{\mG}{\mathcal{G}}

\newcommand{\bT}{\mathbf{T}}

\newcommand{\mL}{\mathcal{L}}

\def\pb #1{\left\{#1\right\}}

\begin{document}
\section{Introduction and Summary}\label{first}
The conformal-traceless decomposition of the gravitational field was
firstly performed in \cite{York:1998hy} in its initial value problem
\footnote{For review and extensive list of references, see
\cite{Gourgoulhon:2007ue}.}. The conformal traceless decomposition
is defined by
\begin{equation}\label{defcon}
h_{ab}=\phi^4 g_{ab} \ , \quad  K_{ab}=\phi^{-2}A_{ab}+\frac{1}{3}
g_{ab}\tau \ ,
\end{equation}
where $h_{ab}$ is spatial physical metric and where $K_{ab}$ is
physical extrinsic curvature. We see that this definition is
redundant since the multiple of the fields $g_{ab},\phi,A_{ab},\tau$
give the same physical metric $g_{ab}$ and extrinsic curvature
$K_{ab}$. In fact, we see that the decomposition (\ref{defcon}) is
invariant under the conformal transformation
\begin{eqnarray}\label{gaugecon}
g'_{ab}(\bx,t)&=&\Omega^4(\bx,t)g_{ab}(\bx,t) \ , \quad
\phi'(\bx,t)=\Omega^{-1}(\bx,t)
\phi(\bx,t) \ , \nonumber \\
A'_{ab}(\bx,t)&=&\Omega^{-2}(\bx,t)A_{ab}(\bx,t), \quad
\tau'(\bx,t)=\tau(\bx,t) \ ,
\nonumber \\
\end{eqnarray}
where $\bx=(x^a, a=1,2,3)$. We also see that (\ref{defcon}) is
invariant under following transformation
\begin{equation}\label{secondsym}
\tau'(\bx,t)=\tau(\bx)+\zeta(\bx,t) \ , \quad A'_{ab}(\bx,t)=
A_{ab}(\bx,t) -\frac{1}{3}\zeta(\bx,t) \phi^6 g_{ab}(\bx,t) \ .
\end{equation}
Clearly the gauge fixing of these symmetries we can eliminate $\tau$
and $\phi$. It is important to stress that in many recent
applications the conformal invariance is broken by imposing the
condition $\sqrt{g}=1$. Further, the second symmetry
(\ref{secondsym}) can be eliminated by imposing the traceless
condition $A_{ab}g^{ab}=0$. Then the variable $\tau$ is the trace of
the extrinsic curvature $K=K_{ab}h^{ab}$. In what follows we do not
impose neither from these conditions.

In this short note we perform the explicit Hamiltonian analysis of
the conformal traceless decomposition of the gravitational field
given in  (\ref{defcon}) following \cite{Brown:2005aq}. We start
with the General Relativity action where the dynamical field is the
physical metric $h_{ab}$. Then we find corresponding Hamiltonian and
then express the action in the Hamiltonian form. As the next step we
introduce  the conformal traceless decomposition (\ref{defcon}) into
this  action, identify the canonical variables and conjugate momenta
and determine corresponding Hamiltonian and spatial
 diffeomorphism constraints. We also identify additional primary
constraint that is a consequence of the symmetry of the theory under
the transformation (\ref{gaugecon}). As a result we find the
Hamiltonian formulation of the General Relativity action that has an
additional two degrees of freedom $\phi$ and its conjugate momenta
$p_\phi$ together with an additional first class constraint.

As the next step we proceed to the analysis of  the gauge fixing of
the constraint that generates (\ref{gaugecon}). Clearly  imposing
the gauge $\phi=0$ we derive the standard General Relativity
Hamiltonian. On the other hand when we introduce  the gauge fixing
function $\mG: \sqrt{g}-1=0$ we obtain theory where the physical
degrees of freedom are the traceless components of the momenta
together with the metric components that obey $\sqrt{g}=1$. We also
have the scalar and corresponding conjugate momenta that measures
the scale factor of the metric. Finally the gauge fixing gives
non-trivial Dirac brackets between these physical variables.

We mean that the gauge fixed form of the theory could be very useful
for the formulation of some alternative versions of the theories of
gravity. In particular we mean that it could be useful for the
formulation of the Ho\v{r}ava-Lifshitz theory when we impose
additional constraint on the scalar mode as in case of
Non-Relativistic Covariant Ho\v{r}ava-Lifshitz gravity
\cite{Horava:2010zj} or as in case of Lagrange multiplier modified
Ho\v{r}ava-Lifshitz gravity \cite{Kluson:2011xx}. It could be also
useful for the formulation of the Spatially Covariant Theories of a
Transverse, Traceless Graviton \cite{Khoury:2011ay}. We hope that we
proceed to the  application of the formalism given  in this paper
for these specific problems in near future.

The structure of this note  is as follows. In the next section
(\ref{second}) we find the Hamiltonian formalism for the conformal
 decomposition of gravity. We identify the primary and secondary
 constraints and calculate the Poisson brackets between them.
 Then in section (\ref{third}) we perform the gauge fixing of the
conformal symmetry and we find corresponding Hamiltonian and
 determine the Dirac brackets between phase space variables.

\section{Hamiltonian Formalism for Conformal
Decomposition of the Gravitational Field}\label{second}
Let us consider four dimensional action for General Relativity
\begin{equation}
S=\frac{1}{2\kappa^2}\int d^4 x\sqrt{-\hg}{}^{(4)}R(\hg) \ ,
\end{equation}
where ${}^{(4)}R$ is four dimensional scalar curvature evaluated on
the four dimensional metric $\hg_{\mu\nu}$. As the next step we
perform $3+1$ decomposition of the metric
\cite{Arnowitt:1962hi,Gourgoulhon:2007ue}.
 Explicitly, we have following relation between four
dimensional metric components $\hg_{\mu\nu}$ and corresponding
spatial metric components $h_{ab}$ and the lapse $N$ and shift
functions $N_a$
\begin{eqnarray}
\hat{g}_{00}=-N^2+N_a h^{ab}N_b \ , \quad \hat{g}_{0a}=N_a \ , \quad
\hat{g}_{ab}=h_{ab} \ ,
\nonumber \\
\hat{g}^{00}=-\frac{1}{N^2} \ , \quad \hat{g}^{0a}=\frac{N^a}{N^2} \
, \quad \hat{g}^{ab}=h^{ab}-\frac{N^a N^b}{N^2} \ .
\nonumber \\
\end{eqnarray}
Note also that $4-$dimensional scalar curvature has following
decomposition
\begin{equation}\label{Rdecom}
{}^{(4)}R=K_{ab}\mG^{abcd}K_{cd}+R \ ,
\end{equation}
where $R$ is three-dimensional spatial curvature, $K_{ab}$ is
extrinsic curvature defined as
\begin{equation}
K_{ab}=\frac{1}{2N} (\partial_t h_{ab} -\nabla_a N_b-\nabla_b N_a) \
,
\end{equation}
where $\nabla_a$ is covariant derivative built from the metric
components $h_{ab}$. Finally we also ignored the boundary terms that
are presented in (\ref{Rdecom}).

Using this formalism we derive the
 General Relativity action in the form
\begin{equation}\label{GRaction}
S=\frac{1}{2\kappa^2}\int d^3\bx dt \sqrt{h}N
(K_{ab}\mG^{abcd}K_{cd}+R) \ ,
\end{equation}
where  $\mG^{ijkl}$ is de Witt metric defined as
\begin{equation}
\mG^{abcd}=\frac{1}{2}(h^{ac}h^{bd}+h^{ad}h^{bc})-h^{ab} h^{cd} \
\end{equation}
with inverse
\begin{equation}
\mG_{abcd}=\frac{1}{2}
(h_{ac}h_{bd}+h_{ad}h_{bc})-\frac{1}{2}h_{ab}h_{cd} \ ,  \quad
\mG_{abcd}\mG^{cdmn}=\frac{1}{2}(\delta_a^m\delta_b^n+
\delta_a^n\delta_b^m) \ .
\end{equation}
However for further purposes we introduce the generalized form of
the de Witt metric that depends on the parameter $\lambda$ as
\begin{equation}
\mG^{abcd}=\frac{1}{2}(h^{ac}h^{bd}+h^{ad}h^{bc})-\lambda
h^{ab}h^{cd} \ , \quad \mG_{abcd}=\frac{1}{2}(h_{ac}h_{bd}+
h_{ad}h_{bc})-\frac{\lambda}{3\lambda-1} h_{ab}h_{cd} \ .
\end{equation}
This generalized  de Witt metric could be useful for the possible
application of given procedure for more general theories of gravity
as for example Ho\v{r}ava-Lifshitz gravity
\cite{Horava:2008ih,Horava:2009uw}.

 In order to perform the Hamiltonian analysis of the conformal
decomposition of the action (\ref{GRaction}) we firstly
 rewrite the
action (\ref{GRaction}) its Hamiltonian form. To do this we
introduce the conjugate momenta
\begin{eqnarray}\label{defmom}
P^{ab}=\frac{\delta S}{\delta
\partial_t h_{ab}}=
\frac{1}{2\kappa^2}\sqrt{h}\mG^{abcd}K_{cd}
\ , \quad P_N=\frac{\delta S}{\delta
\partial_t N}=0 \ , \quad
P_a=\frac{\delta S}{\delta \partial_t N^a}=
0 \ .\nonumber \\
\end{eqnarray}
Then we easily determine corresponding
 Hamiltonian
\begin{equation}
H=\int d^3\bx (\partial_t h_{ab}P^{ab}-\mL)= \int d^3\bx
(N\mH'_T+N^a\mH'_a) \ ,
\end{equation}
where
\begin{equation}
\mH'_T=\frac{2\kappa^2}{\sqrt{h}}
P^{ab}\mG_{abcd}P^{cd}-\frac{1}{2\kappa^2}\sqrt{h}R \ , \quad
\mH'_a=-2h_{ab}\nabla_c\pi^{cb} \ .
\end{equation}
Using the Hamiltonian and the corresponding canonical variables we
write the action (\ref{GRaction}) as
\begin{equation} S=\int dt L=
\int dt d^3 \bx (P^{ab}\partial_t h_{ab}-\mH)= \int dt d^3\bx
(P^{ab}\partial_t h_{ab}-N \mH'_T- N^a\mH'_a) \ .
\end{equation}
Then we insert the decomposition (\ref{defcon}) into the definition
of the canonical momenta $P^{ab}$
\begin{equation}\label{Pabh}
P^{ab}=\frac{1}{2\kappa^2}
\sqrt{g}(\phi^{-4}\tmG^{abcd}A_{cd}
+\frac{1}{3}\phi^2\tau
\tmG^{abcd}g_{cd})
\end{equation}
where the metric $\tmG^{abcd}$ is defined as
\begin{equation}
\tmG^{abcd}=\frac{1}{2}(g^{ac}g^{bd}+g^{ad}g^{bc})-\lambda
g^{ab}g^{cd} \ , \quad \mG^{abcd}=\phi^{-8}\tmG^{abcd} \ .
\end{equation}
Note that $\tmG^{abcd}$ has the inverse
\begin{equation}
\tmG_{abcd}=\frac{1}{2}(g_{ac}g_{bd}+
g_{ad}g_{bc})-\frac{\lambda}{3\lambda-1} g_{ab}g_{cd} \ , \quad
\tmG_{abcd}= \phi^8 \mG_{abcd} \ .
\end{equation}
Using (\ref{Pabh}) and (\ref{defcon}) we rewrite $P^{ab}\partial_t
h_{ab}$ into the form
\begin{eqnarray}
P^{ab}\partial_t h_{ab}&=&
\left(\frac{1}{2\kappa^2}\sqrt{g}\tmG^{abcd}A_{cd} +
\frac{\sqrt{g}}{6\kappa^2} \phi^6(1-3\lambda)\tau
g^{ab}\right)\partial_t g_{ba}+ \nonumber \\
&+& \left(\frac{2}{\kappa^2}\sqrt{g}\phi^{-1}
A_{ab}g^{ba}(1-3\lambda)+\frac{2\sqrt{g}}{\kappa^2}
(1-3\lambda)\phi^5\tau\right)\partial_t\phi
\ .  \nonumber \\
\end{eqnarray}
We see that it is natural to identify the expression in the
parenthesis with  momentum $\pi^{ab}$ conjugate to $g_{ab}$  and
$p_\phi$ conjugate to $\phi$ respectively
\begin{eqnarray}\label{canmom}
\pi^{ab}&=&\frac{1}{2\kappa^2}
\sqrt{g}\tmG^{abcd}A_{cd}+\frac{\sqrt{g}}{6\kappa^2}(1-3\lambda)
\phi^6\tau g^{ab} \ , \nonumber \\
p_\phi&=&\frac{2}{\kappa^2} \sqrt{g}\phi^{-1}
A_{ab}g^{ba}(1-3\lambda)+ \frac{2\sqrt{g}}{\kappa^2}
(1-3\lambda)\phi^5 \tau \ . \nonumber
\\
\end{eqnarray}
Note that we do not impose the traceless condition $g_{ab}A^{ab}=0$.
However using (\ref{canmom}) we can eliminate $\tau$ and determine
following primary constraint
\begin{equation}
\Sigma_D: p_\phi \phi-4\pi^{ab}g_{ba}=0 \ .
\end{equation}
As we will see below this is the constraint that generates conformal
transformation of the dynamical fields. Further,  using
(\ref{canmom}) we find the relation between $P^{ab}$ and $\pi^{ab}$
in the form \footnote{It is important  to  stress that we could
proceed exactly as in \cite{Brown:2005aq} and consider
decomposition of the expression $P^{ab}\partial_t h_{ab}$ in the
form
\begin{eqnarray}\label{exalt}
P^{ab}\partial_t h_{ab}
&=&\frac{2}{\kappa^2}\sqrt{g}\phi^{-1}
A_{ab}g^{ba}(1-3\lambda)\partial_t\phi
+\frac{1}{2\kappa^2}\sqrt{g}\tmG^{abcd}A_{ab}\partial_t
g_{cd}+ \nonumber \\
&+&\frac{2\sqrt{g}}{\kappa^2} (1-3\lambda)\phi^5\tau\partial_t\phi+
\frac{1}{3\kappa^2} \phi^6(1-3\lambda)\tau
\partial_t \sqrt{g} \ .
\nonumber \\
\end{eqnarray}
From (\ref{exalt})  we see   that it is natural to introduce an
additional dynamical variable  $Q$ defined as $Q=\sqrt{ g}$ and
identify corresponding conjugate momenta
\begin{eqnarray}
p_\phi&=&\frac{2}{\kappa^2} \sqrt{g}\phi^{-1}
A_{ab}g^{ba}(1-3\lambda)+ \frac{2\sqrt{g}}{\kappa^2}
(1-3\lambda)\phi^5 \tau \ , \nonumber
\\
\pi^{ab}&=&\frac{1}{2\kappa^2} \sqrt{g}\tmG^{abcd}A_{cd} \ , \quad
p_Q= \frac{1}{3\kappa^2}(1-3\lambda)
\phi^6\tau  \ . \nonumber \\
\end{eqnarray}
 However when we introduce $Q$ as an independent dynamical variable
 we should impose an additional primary constraint $Q-\sqrt{g}=0$.
On the other hand we mean that an existence of the additional
primary constraint would make the analysis more complicated without
any apparent advantages. For that reason we prefer  the
decomposition
 when we have  dynamical variables $g_{ab},\phi$ and
corresponding conjugate momenta $\pi^{ab},p_\phi$ respectively.}
\begin{equation}
P^{ab}=\phi^{-4}\pi^{ab} \ .
\end{equation}
Then we find that the kinetic term in the Hamiltonian constraint
$\mH'_T$ takes the form
\begin{equation} \frac{2\kappa^2}{\sqrt{h}}
P^{ab}\mG_{abcd}P^{cd}= \frac{2\kappa^2\phi^{-6}}{\sqrt{g}}
\pi^{ab}\tmG_{abcd}\pi^{cd} \ .
\end{equation}
 As the next step we introduce the
decomposition (\ref{defcon}) into the contribution $ \int d^3\bx N^a
\mH_a$. Using the  relation between Levi-Civita connections
 evaluated with the metric components $h_{ab}$ and
$g_{ab}$
\begin{equation}
\Gamma_{ab}^c(h)=\Gamma_{ab}^c(g)+ 2\frac{1}{\phi} (\partial_a\phi
\delta^c_b+\partial_b \phi\delta_a^c-\partial_d\phi g^{cd}g_{ab}) \
\end{equation}
and also if we define
 $n_a$ through the relation
 $N_a=\phi^4n_a$
we obtain
\begin{eqnarray}
\int d^3\bx N^a\mH'_a&=&
\int d^3\bx n^a \mH''_a \ , \nonumber \\
\end{eqnarray}
where
\begin{equation}
 \mH''_a=-2g_{ad}D_b \pi^{bd}+
4\phi^{-1}\partial_a \phi g_{cd}\pi^{cd} \ ,
 \nonumber \\
\end{equation}
where the covariant derivative $D_a$ is defined using the
Levi-Civita  connection $\Gamma^c_{ab}(g)$.
Observe that with the help of the constraint $\Sigma_D $
 we can write the constraint $\mH''_a$ as
\begin{eqnarray}
\mH''_a=
 -2g_{ac}D_b \pi^{bc}+ \partial_b\phi p_\phi-
4\phi^{-1}\partial_a\phi \Sigma_D\equiv  \hat{\mH}_a -
4\phi^{-1}\partial_a\phi \Sigma_D
\end{eqnarray}
so that we see that it is natural to
 identify $\hat{\mH}_a$ as an independent constraint. In
 fact, we will see that the smeared form of this constraint
generates the spatial diffeomorphism.

Finally we  proceed to the spatial curvature $R$. Note that there is
a well known relation between $R[h]$ evaluated on $h$ and $R[g]$
evaluated on $g$ so that  we find
\begin{eqnarray}
-\frac{\sqrt{h}}{2\kappa^2}R[h]
=-\frac{\sqrt{g}}{2\kappa^2} \phi^{2}R[g]
+\frac{4\phi\sqrt{g}}{\kappa^2} g^{ab}D_a D_b\phi \ .  \nonumber \\
\end{eqnarray}
Collecting all these results together we obtain  the Hamiltonian
constraint in the form
\begin{equation} \mH'_T=
\frac{2\kappa^2\phi^{-6}}{\sqrt{g}} \pi^{ab}\tmG_{abcd}\pi^{cd}
-\frac{\sqrt{g}}{2\kappa^2} \phi^{2}R +\frac{4\sqrt{g}}{\kappa^2}
\phi  g^{ab}D_a  D_b\phi \
\end{equation}
so that  the action takes the form
\begin{eqnarray}\label{actiontraceless}
S=\int dt d^3\bx(\pi^{ab}\partial_t g_{ab}+ p_\phi\partial_t\phi
-n^a\hat{\mH}_a-N\mH'_T-\lambda \Sigma_D) \ ,
\nonumber \\
\end{eqnarray}
where we included the primary constraint $\Sigma_D$ multiplied by
the Lagrange multiplier $\lambda$.

Now we can proceed to the Hamiltonian analysis of the conformal
decomposition of the gravitational field given  by the  action
(\ref{actiontraceless}). Clearly we have following primary
constraints
\begin{equation}
\pi_N\approx 0 \ , \quad  \pi_a\approx 0 \ , \quad  \Sigma_D\approx
0 \ ,
\end{equation}
where $\pi_N,\pi_a$ are momenta conjugate to $N,n^a$
 with following
non-zero Poisson brackets
\begin{equation}
\pb{N(\bx),\pi_N(\by)}= \delta(\bx-\by) \ , \quad
\pb{n^a(\bx),\pi_b(\by)}= \delta^a_b\delta(\bx-\by) \ .
\end{equation}
Further, the preservation of the primary constraints $\pi_N,\pi_a$
implies following secondary ones
\begin{equation}
\hat{\mH}_a\approx 0 \ , \quad  \mH'_T\approx 0 \ .
\end{equation}
Now we should analyze the requirement of the preservation of the
primary constraint $\Sigma_D$ during the time evolution of the
system. First of all the explicit calculations give
\begin{eqnarray}\label{SigmaDg}
\pb{\Sigma_D(\bx),g_{ab}(\by)}&=&
4g_{ab}(\bx)\delta(\bx-\by) \ , \nonumber \\
\pb{\Sigma_D(\bx),\pi^{ab}(\by)}&=& -4\pi^{ab}(\bx)\delta(\bx-\by) \
, \nonumber
\\
\pb{\Sigma_D(\bx),\phi(\by)}&=& -\phi(\bx)\delta(\bx-\by) \ ,
\nonumber
\\
\pb{\Sigma_D(\bx),p_\phi(\by)}&=& \phi(\bx)\delta(\bx-\by) \
\nonumber
\\
\end{eqnarray}
using the canonical Poisson brackets
\begin{equation}
\pb{g_{ab}(\bx),\pi^{cd}(\by)}= \frac{1}{2} (\delta_a^c\delta_b^d+
\delta_a^d\delta_b^c)\delta(\bx-\by) \ , \quad
\pb{\phi(\bx),p_\phi(\by)}=\delta(\bx-\by) \ .
\end{equation}
 It turns out that it is useful
to introduce the smeared forms of the constraints
$\mH'_T,\hat{\mH}_a,\Sigma_D$
\begin{equation}
\bT_T(N)=\int d^3\bx N \mH'_T \ , \quad \bT_S(N_a)= \int d^3\bx N^a
\hat{\mH}_a \ , \quad \bD(M)=\int d^3\bx M \Sigma_D \ ,
\end{equation}
where $N,N^a$ and $M$ are smooth functions on $\mathbf{R}^3$.  Then
using (\ref{SigmaDg}) and also
\begin{eqnarray}
\pb{\Sigma_D(\bx),\Gamma_{ab}^c(\by)}=2
\delta_b^c\partial_{y^a}\delta(\bx-\by)+2\delta_a^c\partial_{y^b}
\delta(\bx-\by)-2g^{cd}(\by)\partial_{y^d}\delta(\bx-\by)g_{ab}
(\by) \
\nonumber \\
\end{eqnarray}
we easily find that
\begin{equation}\label{bDmHT}
\pb{\bD(M),\mH'_T(\by)}=0 \ .
\end{equation}
To proceed further we use following Poisson brackets
\begin{eqnarray}\label{pbbtSgp}
\pb{\bT_S(N^a),g_{ab}(\bx)}&=&-N^c\partial_c g_{ab}(\bx)-
\partial_a N^cg_{cb}(\bx)-g_{ac}\partial_b N^c(\bx)  \ , \nonumber \\
\pb{\bT_S(N^a),\pi^{ab}(\bx)}&=&-\partial_c (N^c\pi^{ab})(\bx)
+
\partial_c N^a\pi^{cb}(\bx)+\pi^{ac}\partial_c N^b(\bx) \ ,  \nonumber \\
\pb{\bT_S(N^a),\phi(\bx)}&=&-N^a\partial_a\phi(\bx) \ , \nonumber \\
\pb{\bT_S(N^a),p_\phi(\bx)}&=&-\partial_a (N^a p_\phi)(\bx) \ .
\nonumber \\
\end{eqnarray}
Then we easily find
\begin{equation}
\pb{\bT_S(N^a),\Sigma_D(\bx)}=-N^a\partial_a \Sigma_D(\bx)-
\partial_a N^a\Sigma_D(\bx) \
\end{equation}
that together with (\ref{bDmHT}) implies that $\Sigma_D \approx 0$
is the first class constraint.

Now we proceed to the analysis of the preservation of the secondary
constraints $\mH'_T\approx 0 \ , \hat{\mH}_a\approx 0$.  In case of
$\hat{\mH}_a$ we find following Poisson brackets
\begin{equation}
\pb{\hat{\mH}_a(\bx),\hat{\mH}_b(\by)}=
\hat{\mH}_b(\bx)\frac{\partial} {\partial
x^a}\delta(\bx-\by)-\hat{\mH}_a(\by)\frac{\partial}{\partial y^b}
\delta(\bx-\by) \
\end{equation}
which implies that the smeared form of the diffeomorphism
constraints takes the familiar form
\begin{equation}
\pb{\bT_S(N^a),\bT_S(M^b)}= \bT_S(N^b\partial_b M^a- M^b\partial_b
N^a) \ .
\end{equation}
Further using (\ref{pbbtSgp}) we easily find
\begin{equation}
\pb{\bT_S(N^a),\mH'_T(\bx)}= -\partial_c N^c\mH'_T(\bx) -
N^c\partial_c \mH'_T(\bx)
\end{equation}
or equivalently
\begin{equation}
\pb{\bT_S(N^a),\bT_T(M)}=\bT_T(N^a\partial_a M) \ .
\end{equation}
These results show that $\hat{\mH}_a$ are the first class
constraints.

 Finally we have to calculate the Poisson brackets between
Hamiltonian constraints. To do this we use the fact that
\begin{equation}
\pb{R(\bx),\pi^{ab}(\by)}= -R^{ab}(\bx)
\delta(\bx-\by)+D^aD^b\delta(\bx-\by)-g^{ab} D_cD^c\delta(\bx-\by) \
.
\end{equation}
Then after some lengthy calculations we derive following Poisson
bracket
\begin{eqnarray}\label{bTT(NM)}
\pb{\bT_T(N),\bT_T(M)}&=& -\frac{8}{3\lambda-1} \int d^3\bx
(\partial_h N M-\partial_h M N) \phi^{-5} g^{hp}
\partial_p\phi g_{cd}\pi^{cd}+\nonumber \\
&+&2\int d^3\bx (\partial_d NM-\partial_d M N)\phi^{-4} D_c\pi^{cd}+
\nonumber \\
&+&2\frac{\lambda-1} {3\lambda-1}\int d^3\bx (\partial_h N M
-\partial_h M N) g^{hm}D_m (g_{cd}\pi^{cd}) \ .  \nonumber \\
\end{eqnarray}
First of all we see that in order to eliminate the last term we have
to demand that the parameter $\lambda$ is equal to one. In what
follows we will then presume that $\lambda=1$ keeping in mind that
the general case of $\lambda\neq 1$ could be useful when we perform
the conformal decomposition of the metric in case of
Ho\v{r}ava-Lifshitz gravity. For $\lambda=1$ we obtain that
(\ref{bTT(NM)}) can be written as
\begin{eqnarray}
\pb{\bT_T(N),\bT_T(M)}&=& \nonumber \\
&=&\bT_S((\partial_b MN-\partial_b N M)g^{ba}\phi^{-4})+ \bD(
(\partial_a N M-\partial_a M N) \phi^{-5} g^{ab}
\partial_b\phi) \ .  \nonumber \\
\end{eqnarray}
This result implies that the Poisson bracket between Hamiltonian
constraints vanishes on the constraint surface.

Now we can outline our results. We performed  the Hamiltonian
analysis of the action (\ref{actiontraceless}) and we identified the
first class constraints $\pi_N\approx 0 \ , \pi_a\approx 0 \ ,
\mH'_T\approx 0 \ , \hat{\mH}_a\approx 0$ and $\Sigma_D\approx 0$.
On the other hand we  have following phase space degrees of freedom
$N,\pi_N,n^a,\pi_a,g_{ab},\pi^{cd}$ and $\phi,p_\phi$. Then using
the standard counting of the physical degrees of freedom
\cite{Henneaux:1992ig}  we find that given theory has four physical
degrees of freedom which is the correct number of the degrees of
freedom of the General Relativity.

\section{Fixing  Gauge Symmetry $\Sigma_D\approx 0$}\label{third}
We saw in previous section that conformal decomposition of the
gravitational field implies an existence of the additional scalar
field $\phi$ together with the first class constraint
$\Sigma_D\approx 0$ that generates the conformal transformation.
  The simplest way  how to fix given symmetry
  is to impose  the constraint $\phi=0$ which however leads  to the
standard General Relativity Hamiltonian. Clearly this is not very
interesting result. For that reason we  rather  consider following
form of the gauge fixing function
\begin{equation}\label{mG}
\mG(\bx): \sqrt{g}(\bx)-1=0 \ .
\end{equation}
In this case the scalar $\phi$ has the physical meaning as the scale
factor of the metric. Now we would like to see the consequence of
the gauge fixing (\ref{mG}) for the structure of the theory.

As the first step we should note that  the extended Hamiltonian now
contains the constraint $\mG$ which,  in order to  fix the gauge has
to have non-zero Poisson brackets with $\Sigma_D$ and also $\mG$ has
to be preserved during the time evolution of the system. In fact, we
have to check that all constraints are now preserved when the
extended Hamiltonian contains the additional constraint $\mG\approx
0$. Explicitly
\begin{equation}\label{HT}
H_T=\int d^3\bx \left(N\mH'_T+n^a\hat{\mH}_a+ v^N\pi_N+v^a\pi_a+
\lambda \Sigma_D+\Gamma \mG\right) \ .
\end{equation}
Then using  following Poisson bracket
\begin{eqnarray}
\pb{\sqrt{g}(\bx),\pi^{ab}(\by)}&=&\frac{1}{2} g^{ab}\sqrt{g}(\bx)
\delta(\bx-\by) \  \nonumber \\
\pb{\mH'_T(\by),\mG(\bx)}&=&\kappa^2\phi^{-6}g_{cd}\pi^{cd}(\bx)
\delta(\bx-\by)
 \ , \nonumber \\
 \pb{\Sigma_D(\by),\mG(\bx)}&=&6\sqrt{g}(\bx)\delta(\bx-\by)
 \approx 6\delta(\bx-\by)\equiv \triangle_{\Sigma_D,\mG}(\bx,\by)
 \nonumber \\
 \end{eqnarray}
we find  equations that determine the time evolution of the
constraints
\begin{eqnarray}
\partial_t \mG(\bx)&=&-\pb{H_T,\mG(\bx)}=\nonumber \\
&=&
-\left(N\kappa^2\phi^{-6}g_{ab}\pi^{ba}(\bx)+6\sqrt{g}\lambda(\bx)
-\partial_a N^a(\bx) -\partial_a N^a\mG(\bx)-N^a\partial_a
\mG(\bx)\right) \ ,
 \nonumber \\
\partial_t\mH'_T(\bx)&=&-\pb{H_T,\mH'_T(\bx)}
\approx -\int d^3\by \Gamma(\by)\pb{\mG(\by),\mH'_T(\bx)}=
\Gamma\kappa^2\phi^{-6}g_{cd}\pi^{cd}(\bx)=0 \ , \nonumber \\
\partial_t\Sigma_D(\bx)&=&-\pb{H_T,\Sigma_D(\bx)}
\approx -\int d^3\by\Gamma(\by)\pb{\mG(\by),\Sigma_D(\bx)}=
6\sqrt{g}
\Gamma(\bx)=0 \ .  \nonumber \\
\end{eqnarray}
We see that the last two equations has the solution $\Gamma=0$ while
the first one gives
\begin{equation}\label{lambda}
\lambda=-\frac{N\kappa^2\phi^{-6}g_{ab}\pi^{ba}}{6\sqrt{g}}+\frac{1}{6\sqrt{g}}\partial_a
N^a \ .
\end{equation}
Inserting (\ref{lambda}) together with $\Gamma=0$ into the extended
Hamiltonian $H_T$ we find
\begin{eqnarray}
H_T
&=& \int d^3\bx  \left(N\left[\mH_T-\frac{\kappa^2
\phi^{-6}}{6\sqrt{g}}g_{cd}\pi^{cd}\Sigma_D\right]+v^N\pi_N+v^a\pi_a\right)+\nonumber \\
&+&\int d^3\bx N^a\left(-2g_{ac}\nabla_d\pi^{cd}-\partial_a
[\frac{1}{6\sqrt{g}}\Sigma_D]+p_\phi \partial_a\phi\right) \equiv
\nonumber \\
&\equiv & \int
d^3\bx ( N \tmH_T+N^a \tmH_a+v^N\pi_N+v^a\pi_a) \ . \nonumber \\
\end{eqnarray}
We claim that the Hamiltonian on the reduced phase space is given as
the linear combinations of the first class constraints
$\tmH_T,\tmH_a$ together with the second class constraints
$\Sigma_D,\mG$.

To see this explicitly we again introduce the smeared form of these
constraints
\begin{eqnarray}
\tbT_T(N)&=& \int d^3\bx \tmH_T= \bT_T(N)-\bD\left(N\frac{\kappa^2
\phi^{-6}}{6\sqrt{g}}g_{cd}\pi^{cd}\right) \ , \nonumber \\
 \tbT_S(N^a)&=&\int d^3\bx
N^a \tmH_a=\bT_S(N^a)+\bD\left(\frac{1}{6\sqrt{g}} \partial_a
N^a\right) \ .
\end{eqnarray}
Now using the Poisson brackets determined in previous section we see
that the Poisson brackets between $\tbT_T,\tbT_S$ are proportional
to the constraints and hence vanish on the constraint surface. It is
also clear that we have
\begin{equation}
\pb{\tbT_T(N),\mG(\bx)}=0 \ , \quad \pb{\tbT_S(N^a),\mG(\bx)}=0
\end{equation}
together with $\pb{\tbT_T(N),\bD(M)}=0 \ ,
\pb{\tbT_S(N^a),\bD(M)}=0$ which show that $\tmH_T,\tmH_a$ are the
first class constraints.

As the next step we have to eliminate the second class constraints
which can be done when we replace the Poisson brackets with
corresponding Dirac brackets.
Explicitly we find
\begin{eqnarray}
\pb{g_{ab}(\bx),\pi^{cd}(\by)}_D&=& \pb{g_{ab}(\bx),\pi^{cd}(\by)}-
\int d\bz d\bz'\pb{g_{ab}(\bx),\Sigma_D(\bz)}
\triangle^{\Sigma_D,\mG}(\bz,\bz')\pb{\mG(\bz'),\pi^{cd}(\by)}- \nonumber \\
&-&\int d\bz d\bz'\pb{g_{ab}(\bx),\mG(\bz)}
\triangle^{\mG,\Sigma_D}(\bz,\bz')\pb{\Sigma_D(\bz'),\pi^{cd}(\by)}= \nonumber \\
&=&\frac{1}{2}(\delta_a^c\delta_b^d+\delta_a^d\delta_b^c)\delta(\bx-\by)-
\frac{1}{3}g_{ab}g^{cd}(\bx)\delta(\bx-\by) \ , \nonumber \\
\end{eqnarray}
where $\triangle^{\mG,\Sigma_D}$ is the inverse to
$\triangle_{\Sigma_D,\mG}$ with following non-zero components
\begin{equation}
\triangle^{\mG,\Sigma_D}(\bx,\by)=\frac{1}{6}\delta(\bx-\by) \ ,
\quad \triangle^{\Sigma_D,\mG}(\bx,\by)=-\frac{1}{6}\delta(\bx-\by)
\ .
\end{equation}
In the same way we find
\begin{eqnarray}
\pb{\pi^{ab}(\bx),\pi^{cd}(\by)}_D&=&
\frac{1}{3}\left(\pi^{ab}g^{cd}-g^{ab}\pi^{cd}\right)(\bx)\delta(\bx-\by)
 \ , \nonumber \\
\pb{\pi^{ab}(\bx),p_\phi(\by)}_D&=&
\frac{1}{6}g^{ab}p_\phi(\bx)\delta(\bx-\by)  \ ,
\pb{\pi^{ab}(\bx),p_\phi(\by)}_D=-\frac{1}{6}g^{ab}p_\phi(\bx)\delta(\bx-\by)
\ .
\nonumber \\
\end{eqnarray}
If we define $\pi=\pi^{ab}g_{ba}$ we find
\begin{eqnarray}
\pb{g_{ab}(\bx),\pi(\by)}_D=0 \ , \quad
\pb{\pi^{ab}(\bx),\pi(\by)}_D=0 \ . \nonumber \\
\end{eqnarray}
It turns out that it is useful to introduce the traceless part of
the conjugate momentum $\tpi^{ab}$
\begin{equation}
\tilde{\pi}^{ab}=\pi^{ab}-\frac{1}{3}g^{ab}\pi  \ , \quad
\tpi^{ab}g_{ba}=0 \ .
\end{equation}
Using previous results we derive following  Dirac brackets
\begin{eqnarray}
\pb{g_{ab}(\bx),\tpi^{cd}(\by)}_D&=&
\frac{1}{2}(\delta_a^c\delta_b^d+\delta_a^d\delta_b^c)\delta(\bx-\by)-
\frac{1}{3}g_{ab}g^{cd}(\bx)\delta(\bx-\by) \ , \nonumber \\
 \pb{\tpi^{ab}(\bx),\tpi^{cd}(\by)}_D&=&
\frac{1}{3}\left(\pi^{ab}g^{cd}-g^{ab}\pi^{cd}\right)(\bx)\delta(\bx-\by)
\ ,  \nonumber \\
 \pb{\tpi^{ab}(\bx),p_\phi(\by)}_D&=&0 \ ,
 \quad \pb{\tpi^{ab}(\bx),\phi(\by)}_D=0 \
 \ . \nonumber \\
\end{eqnarray}
%
We showed that $\tmH_T$ together with $\tmH_a$ are
 the first class constraints. We also identified the second
class constraints $\Sigma_D,\mG$. According to the standard analysis
of the constraint systems  these constraints can be solved
explicitly. Solving the constraint $\Sigma_D=0$ for $\pi$ we obtain
 $\pi_{ab}g^{ab}=\frac{1}{4}p_\phi\phi$ while solving the constraint
 $\mG$ gives  $\sqrt{g}=1$.
Then the Hamiltonian constraint $\tmH_T$ takes the form
\begin{eqnarray}
\tmH_T=2\kappa^2\phi^{-6}
\tilde{\pi}^{ab}g_{ac}g_{bd}\tilde{\pi}^{cd} -\frac{1}{2\kappa^2}
\phi^{2}R +\frac{4}{\kappa^2} \phi  g^{ab}D_a
D_b\phi-\frac{\kappa^2}{24}\phi^{-4}p_\phi^2 \ .  \nonumber \\
\end{eqnarray}
Observe that $\tmH_T$ depends on $\tpi^{ab},g_{ab}$ that have $8$
phase space degrees of freedom together with $p_\phi,\phi$.
%
In the same way we find that
 $\tmH_a$ is equal to
\begin{equation}
\tmH_a=-2g_{ac}\nabla_d\tpi^{cd} -\frac{1}{6}\partial_a (p_\phi
\phi)+p_\phi\partial_a\phi \ .
\end{equation}
It is interesting to determine the Dirac bracket between the smeared
form of the constraints $\tmH_a$ and the canonical variables. In
fact, since $\tmH_a$ are the first class constraints we find that
the Dirac brackets between them  and any phase space variable
coincides with corresponding Poisson bracket. Then we obtain
\begin{eqnarray}
\pb{\tbT_S(N^m),g_{ab}(\bx)}_D
&=&-N^c\partial_c g_{ab}(\bx)-
\partial_a N^cg_{cb}(\bx)-g_{ac}\partial_b N^c(\bx)
-\frac{2}{3}\partial_c N^c g_{ab}(\bx) \ , \nonumber \\
\pb{\tbT_S(N^m),\tpi^{ab}(\bx)}_D&=&-N^c\partial_c \tpi^{ab}(\bx)
+\partial_c N^a \tpi^{cb}(\bx)+ \tpi^{ac}\partial_c N^b(\bx)
-\frac{1}{3}\partial_c
N^c\tpi^{ab}(\bx) \ ,  \nonumber \\
\pb{\tbT_S(N^m),\phi(\bx)}_D&=& -N^c\partial_c\phi(\bx)-\frac{1}{6}
\partial_m N^m\phi (\bx) \ , \nonumber \\
\pb{\tbT_S(N^m),p_\phi(\bx)}_D&=& -N^c\partial_c
p_\phi(\bx)-\frac{5}{6}
\partial_c N^c p_\phi (\bx) \  . \nonumber \\
\end{eqnarray}
Then  after some calculations we find
\begin{equation}
\pb{\tbT_S(N^a),\tmH_T(\bx)}= -N^m\partial_m \tmH_T(\bx) -\partial_m
N^m \tmH_T(\bx) \
\end{equation}
which is desired result since it shows that the Hamiltonian
constraint transforms as the tensor density under spatial
diffeomorphism generated by $\tbT_S(N^a)$.

 Let us outline results derived in this section.
We  fix of the conformal symmetry by imposing the condition
$\sqrt{g}=1$. Then we find that the
 dynamical fields $\phi,p_\phi$ together with  $\tpi^{ab},g_{ab}$
where $\sqrt{g}=1$ and where $\tpi^{ab}g_{ba}=0$. We also showed
that given there are four first class constraints $\tmH_T,\tmH_a$.
 Then
 we can proceed further and perform the gauge fixing of
some of these first class constraints.  In fact, we can fix the
Hamiltonian constraint $\tmH_T$ in order to eliminate the scalar
field degrees of freedom $p_\phi,\phi$ so that the reduced phase
space will be governed by $g_{ab},\tpi^{ab}$ with the three first
class constraints $\tmH_a$. Note that due to the presence of these
constraints the number of physical degrees of freedom is four. We
mean  that the theory formulated with $g_{ab},\tpi^{ab}$ which is
invariant under the spatial diffeomorphism could be the starting
point for the alternative formulations of theory of gravity, see for
example \cite{Khoury:2011ay}.

 \noindent {\bf
Acknowledgements:}
 This work   was
supported by the Grant agency of the Czech republic under the grant
P201/12/G028. \vskip 5mm



\end{document}